\documentclass[conference]{IEEEtran}
\IEEEoverridecommandlockouts

\usepackage{cite}
\usepackage{amsmath,amssymb,amsfonts}
\usepackage{algorithmic}
\usepackage{graphicx}
\usepackage{textcomp}
\usepackage{xcolor}
\usepackage{booktabs}
\usepackage{multirow}
\usepackage{url}
\usepackage{microtype}
\usepackage{array}

\graphicspath{{figures/}}

\def\BibTeX{{\rm B\kern-.05em{\sc i\kern-.025em b}\kern-.08em
    T\kern-.1667em\lower.7ex\hbox{E}\kern-.125emX}}

\begin{document}

\title{REAN: Reconstruction-aware ECG Anonymization Based on Privacy--Utility Orthogonality}

\author{
\IEEEauthorblockN{1\textsuperscript{st} Taerin Ki}
\IEEEauthorblockA{\textit{Chung-Ang University} \\
Seoul, South Korea \\
rlxofls@cau.ac.kr}
\and
\IEEEauthorblockN{2\textsuperscript{nd} Sunghwan Park}
\IEEEauthorblockA{\textit{Chung-Ang University} \\
Seoul, South Korea \\
tjdghks994@cau.ac.kr}
\and
\IEEEauthorblockN{3\textsuperscript{rd} Junyoung Park}
\IEEEauthorblockA{\textit{Chung-Ang University} \\
Seoul, South Korea \\
june295921@cau.ac.kr}
\and
\IEEEauthorblockN{4\textsuperscript{th} Jaewoo Lee}
\IEEEauthorblockA{\textit{Chung-Ang University} \\
Seoul, South Korea \\
jaewoolee@cau.ac.kr}
}

\maketitle

\begin{abstract}
A shared electrocardiogram (ECG) is itself a biometric fingerprint that can re-identify a patient and reveal personal information.
Recent ECG anonymizers transform the signal before sharing to reduce privacy leakage.
However, existing methods still face a privacy--utility trade-off, in which preserving privacy often compromises utility while preserving utility reveals personal information.
We propose \emph{REAN} (\emph{RE}construction-aware ECG \emph{AN}onymizer), a raw ECG signal anonymizer, to address this privacy--utility trade-off.
REAN reconstructs the signal using a 1-D U-Net trained with losses from frozen privacy and utility classifiers to reduce privacy leakage while preserving utility.
The privacy and utility gradients are near-orthogonal ($\approx$93.8$^\circ$), so reducing privacy leakage leaves utility almost unchanged.
On four public PhysioNet databases, REAN achieves the strongest privacy--utility balance among raw ECG signal baselines.
It drives re-identification to chance (0.96$\to$0.00), keeps arrhythmia macro-AUROC at the clean level (Clean 0.9982 vs.\ REAN 0.9991), and maintains re-identification protection under unseen privacy-classifier architectures.
\end{abstract}


\section{Introduction}

Electrocardiograms (ECGs) are widely shared for diagnosing arrhythmia, ischemia, and conduction disorders, yet the same waveform is a biometric fingerprint that reveals a patient's identity, gender, and age~\cite{ecgunveiled,transecg}.
Removing identifiers does not protect the patient, because the waveform itself is a quasi-identifier that enables re-identification from partial side knowledge~\cite{linkageattacks,deanon}.
An effective defense must therefore transform the signal itself rather than its metadata.

\begin{figure}[t]
\centering
\includegraphics[width=\columnwidth]{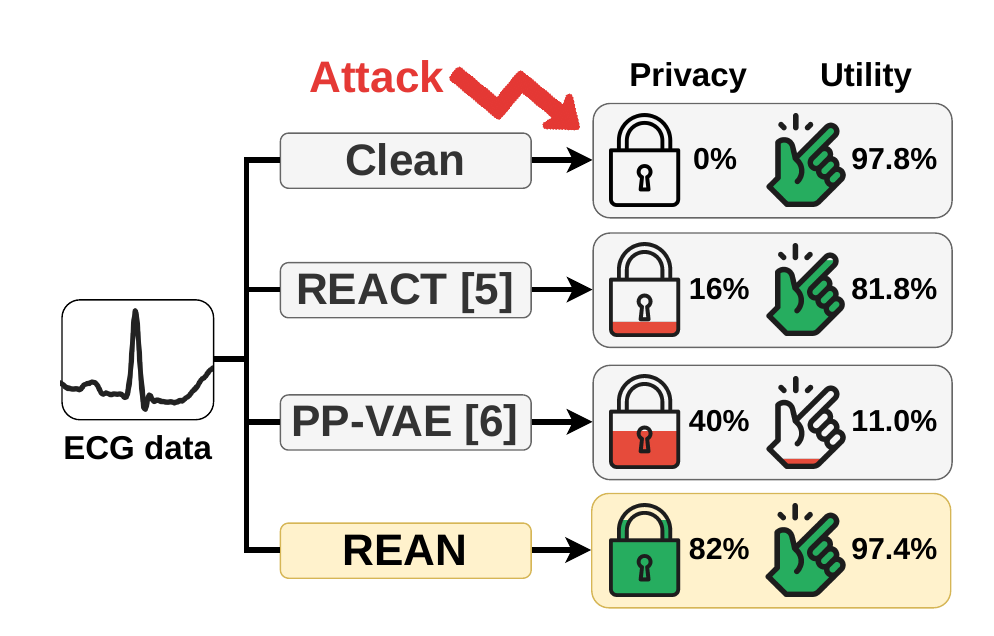}
\caption{The privacy--utility trade-off across ECG anonymizers. The lock marks how much identity and personal-information leakage the defense removes, and the hand marks the arrhythmia-diagnosis accuracy retained after anonymization, combined so that leaving any single attribute exposed keeps the score low. Only REAN scores high (green) on both axes.}
\label{fig:teaser}
\end{figure}

Transforming the signal faces a privacy--utility trade-off between hiding biometric information and preserving diagnosis.
Recent signal-level methods either degrade diagnosis or leave biometric information recoverable~\cite{react,ppvae,privecg,leeetal}, because the two are entangled in the same morphology.
This trade-off is usually treated as the unavoidable price of anonymization.
Fig.~\ref{fig:teaser} shows this trade-off across existing anonymizers.
We instead ask the central question of this paper.
\emph{Can biometric information be removed from an ECG without damaging the morphology that diagnosis depends on?}

We answer this question by observing that the utility and privacy directions are nearly orthogonal in ECG signal space.
Measured as input gradients, the two directions meet at about $89.8^\circ$ (Section~\ref{sec:ortho}).
Biometric information can therefore be suppressed along the privacy direction with almost no effect on diagnosis.

We propose REAN (\emph{RE}construction-aware ECG \emph{AN}onymizer), a 1-D U-Net that exploits this geometry and anonymizes an ECG in a single forward pass.
REAN trains one objective that preserves diagnosis, protects privacy, and limits distortion, using a frozen diagnostic classifier and three frozen biometric classifiers as training signals.
The utility and privacy terms optimize together with little conflict, because their gradients are orthogonal.
A learned residual conditioned on the input follows the privacy direction and hides biometric information at little cost to diagnosis, whereas blind noise moves the signal in every direction and harms diagnosis.

This paper makes the following contributions.
\begin{itemize}
\item \textbf{Privacy--utility orthogonality.} We show that the utility and privacy directions are nearly orthogonal in ECG signal space, and use this geometry as the basis for anonymization (Section~\ref{sec:ortho}).
\item \textbf{The REAN anonymizer.} We propose REAN, a classifier-guided 1-D U-Net that turns this geometry into a single training objective and anonymizes a raw ECG waveform in a single forward pass (Section~\ref{sec:method}).
\item \textbf{Trade-off resolution on PhysioNet databases.} REAN drives re-identification from $0.964$ to chance while keeping arrhythmia macro-AUROC statistically indistinguishable from clean, and stays robust under purification and attackers unseen during training (Section~\ref{sec:main}).
\end{itemize}

\section{Preliminary}
\label{sec:prelim}

\subsection{The ECG signal}
\label{sec:ecg}
\begin{figure}[t]
\centering
\includegraphics[width=0.80\columnwidth]{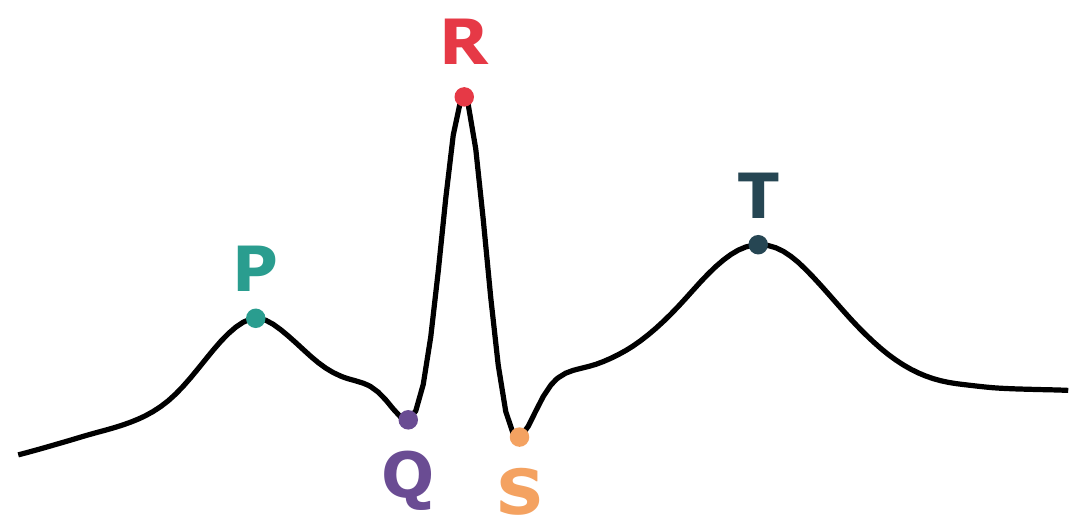}
\caption{ECG waveform for a normal cardiac cycle. Identity is carried mostly by the QRS shape, while arrhythmia diagnosis depends on rhythm and the broader P and T morphology.}
\label{fig:morph}
\end{figure}

An ECG carries both diagnostic and biometric information in the same waveform.
The P--QRS--T morphology and the R--R interval are markers of arrhythmia, and the same shapes are person-specific enough to identify the patient (Fig.~\ref{fig:morph}).
Biometric information rests mostly on the person-specific QRS shape, whereas diagnosis depends on rhythm and the broader P and T morphology, so the two need not occupy the same part of the signal.
We write an ECG as a signal window $\mathbf{x}\in\mathbb{R}^{L}$ and instantiate it at 250\,Hz with 8-second windows ($L{=}2000$).
The property we study does not depend on a specific $L$ or sampling rate.

\subsection{Orthogonality of the utility and privacy directions}
\label{sec:ortho}
The utility and privacy directions are nearly orthogonal in ECG signal space, and this geometry is what makes REAN possible.
We establish it with frozen classifiers alone, before any anonymizer or training objective.

For a signal window $\mathbf{x}\in\mathbb{R}^{L}$, a frozen diagnostic classifier $f_{\mathrm{diag}}$ gives the utility loss $\mathcal{L}_{\mathrm{util}}=\mathrm{CE}(f_{\mathrm{diag}}(\mathbf{x}),y_{\mathrm{diag}})$, and frozen biometric classifiers $f_a$ for the protected attributes $a\in\mathcal{A}=\{\mathrm{reid},\mathrm{gender},\mathrm{age}\}$ give the privacy loss
\begin{equation}
\mathcal{L}_{\mathrm{priv}} = \sum_{a\in\mathcal{A}} \mathrm{CE}\big(f_a(\mathbf{x}),\,y_a\big).
\label{eq:priv}
\end{equation}
The utility and privacy directions are the input gradients $\nabla_{\mathbf{x}}\mathcal{L}_{\mathrm{util}}$ and $\nabla_{\mathbf{x}}\mathcal{L}_{\mathrm{priv}}$, the directions that most change each loss.

\begin{figure}[t]
\centering
\includegraphics[width=\columnwidth]{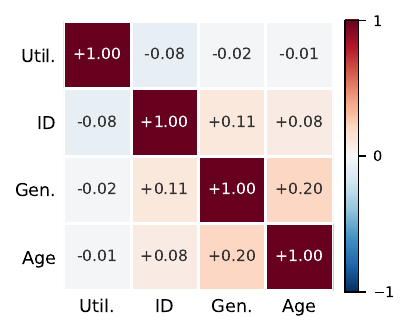}
\caption{Pairwise cosine between the parameter-space gradients of the utility (arrhythmia), ReID, gender, and age losses on clean ECG. Off-diagonals are near zero: the diagnostic gradient is near-orthogonal to every biometric gradient (utility vs.\ privacy $\approx 93.8^\circ$, $\cos\approx-0.066$).}
\label{fig:ortho}
\end{figure}

The two directions are near-orthogonal in signal space.
On clean ECG the per-window angle between them averages about $89.8^\circ$ (cosine $0.003$), and every protected attribute shows the same pattern ($89.6^\circ$--$89.95^\circ$).
The angle stays stable across 4000 windows ($98.5\%$ within $80^\circ$--$100^\circ$) and across biometric backbones (ResNet, InceptionTime, CNN-LSTM), as reported in Table~\ref{tab:robust}.
The same orthogonality holds in the classifiers' parameter space, where $\nabla_\theta\mathcal{L}_{\mathrm{util}}$ and $\nabla_\theta\mathcal{L}_{\mathrm{priv}}$ meet at about $93.8^\circ$ ($\cos\approx-0.066$) and the diagnostic gradient is near-orthogonal to every individual biometric gradient (Fig.~\ref{fig:ortho}).

A near-$90^\circ$ angle is not meaningful on its own, since two random vectors in $\mathbb{R}^{2000}$ are already nearly orthogonal.
What rules out coincidence is the causal consequence.
Full-signal Gaussian noise, which ignores this structure, collapses arrhythmia macro-AUROC to $\approx0.62$, whereas a perturbation along the privacy direction leaves diagnosis at the clean level (Section~\ref{sec:main}).

This geometry sets two requirements for an anonymizer.
An anonymizer should move the signal along the privacy direction, which suppresses biometric information at almost no cost to diagnosis (signal space), and it should optimize privacy and utility together, which their orthogonal gradients permit (parameter space).
REAN meets both.

\begin{table}[t]
\centering
\caption{Robustness of the orthogonality (4000 clean windows, input-gradient angle). The utility--privacy angle stays near $90^\circ$ across protected attributes and across attacker backbones.}
\label{tab:robust}
\setlength{\tabcolsep}{5pt}
\begin{tabular}{lccc}
\toprule
Measurement & Angle & $\cos$ (95\% CI) & 80--100$^\circ$ \\
\midrule
Utility vs.\ privacy (bundled) & $89.8^\circ$ & 0.003 ($\pm$0.002) & 98.5\% \\
Utility vs.\ identity          & $89.8^\circ$ & 0.004 ($\pm$0.002) & 98.0\% \\
Utility vs.\ gender            & $89.95^\circ$ & 0.001 ($\pm$0.002) & 99.0\% \\
Utility vs.\ age               & $89.6^\circ$ & 0.007 ($\pm$0.002) & 98.8\% \\
\;\;backbone: ResNet           & $89.5^\circ$ & 0.008 ($\pm$0.004) & 83.2\% \\
\;\;backbone: InceptionTime    & $89.9^\circ$ & 0.002 ($\pm$0.004) & 86.1\% \\
\;\;backbone: CNN-LSTM         & $89.5^\circ$ & 0.008 ($\pm$0.003) & 92.4\% \\
\bottomrule
\end{tabular}
\end{table}

\section{REAN: Reconstruction-aware ECG Anonymizer}
\label{sec:method}
\begin{figure*}[t]
\centering
\includegraphics[width=\textwidth]{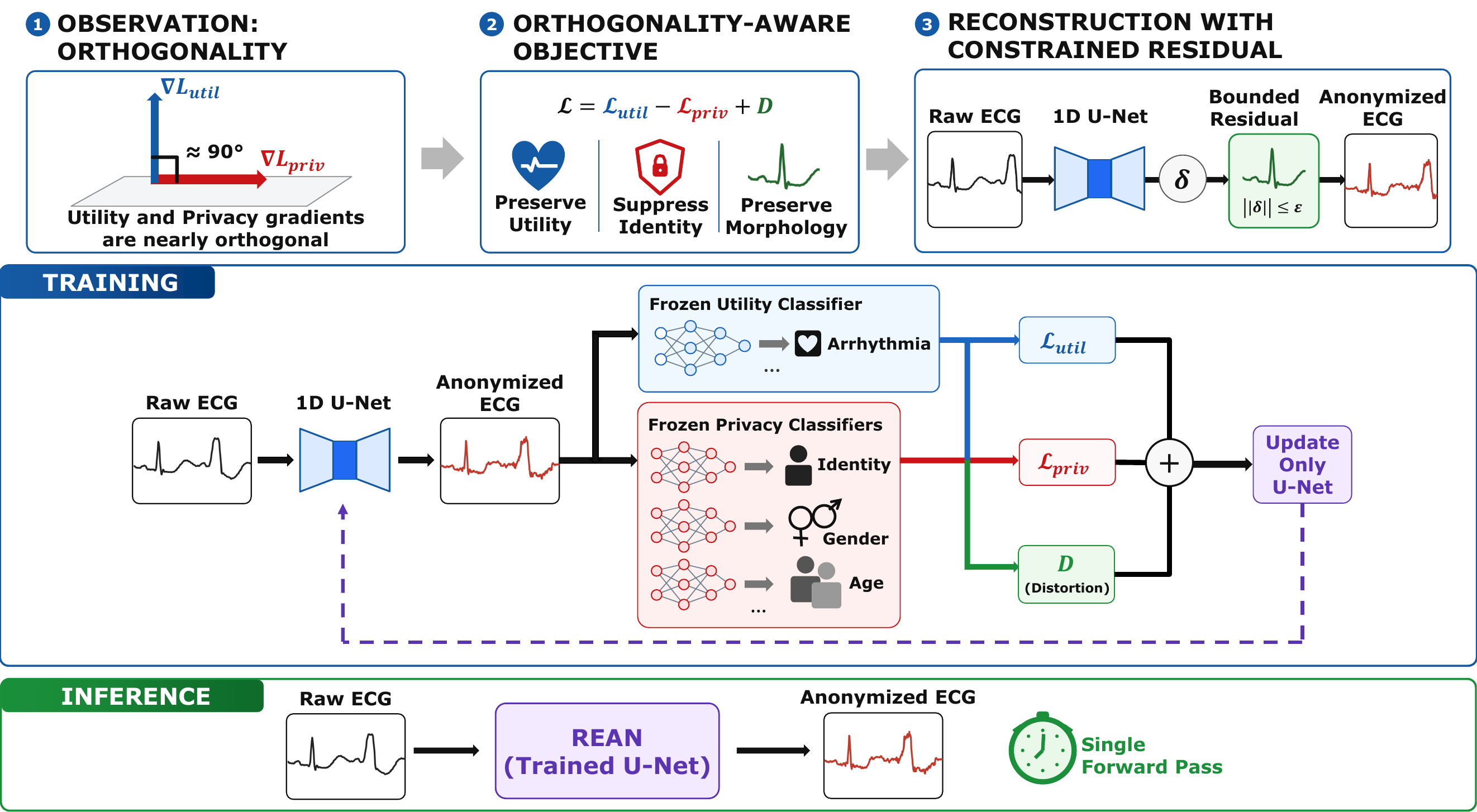}
\caption{REAN framework, shown as three design steps above the training and inference pipelines. (1)~On clean ECG the utility and privacy gradients $\nabla\mathcal{L}_{\mathrm{util}}$ and $\nabla\mathcal{L}_{\mathrm{priv}}$ are nearly orthogonal ($\approx 90^\circ$). (2)~This geometry motivates a single orthogonality-aware objective $\mathcal{L}=\mathcal{L}_{\mathrm{util}}-\mathcal{L}_{\mathrm{priv}}+\mathcal{D}$ that preserves diagnosis, suppresses biometric leakage, and limits distortion. (3)~A 1-D U-Net reconstructs the waveform by adding a bounded, input-conditioned residual (Eq.~\ref{eq:output}), turning a raw ECG into an anonymized ECG. In training, a frozen utility classifier (arrhythmia) and three frozen privacy classifiers (identity, gender, age) score the anonymized output, and only the U-Net is updated. At inference, the trained U-Net anonymizes each window in a single forward pass.}
\label{fig:overview}
\end{figure*}

REAN anonymizes an ECG window in a single forward pass by exploiting the orthogonality of Section~\ref{sec:ortho} (Fig.~\ref{fig:overview}).
We first design the training objective that suppresses biometrics without harming diagnosis (Section~\ref{sec:loss}), then realize it as a 1-D U-Net that reconstructs the waveform along the privacy direction (Section~\ref{sec:unet}).

\subsection{Training objective}
\label{sec:loss}
REAN trains a single objective with three terms that preserve diagnosis, suppress biometric information, and limit distortion.
The frozen diagnostic classifier $f_{\mathrm{diag}}$ scores diagnosis and the three frozen biometric classifiers $f_a$ score biometric leakage; they never change during training and only score a candidate output $\hat{\mathbf{x}}$.
With cross-entropy $\mathrm{CE}(\cdot,\cdot)$, the utility loss is $\mathcal{L}_{\mathrm{util}}=\mathrm{CE}(f_{\mathrm{diag}}(\hat{\mathbf{x}}), y_{\mathrm{diag}})$ and the privacy loss is the bundled $\mathcal{L}_{\mathrm{priv}}=\mathrm{CE}_{\mathrm{bio}}$ of Eq.~\eqref{eq:priv}.
The objective is
\begin{equation}
\mathcal{L} = \mathcal{L}_{\mathrm{util}}
  - \mathcal{L}_{\mathrm{priv}}
  + \mathcal{D}(\mathbf{x},\hat{\mathbf{x}}),
  \label{eq:loss}
\end{equation}
where $\mathcal{D}(\mathbf{x},\hat{\mathbf{x}}) =\mathrm{PRD}(\mathbf{x},\hat{\mathbf{x}})+\|\hat{\mathbf{x}}-\mathbf{x}\|_1$ limits distortion.
The utility and distortion terms are minimized, while the privacy term is maximized through the minus sign.

The three terms can be optimized together because their gradients are near-orthogonal.
The minus sign on $\mathcal{L}_{\mathrm{priv}}$ raises the worry that $\nabla_\theta\mathcal{L}_{\mathrm{util}}$ and $\nabla_\theta\mathcal{L}_{\mathrm{priv}}$ subtract and leave little to train on.
Orthogonality rules this out.
With $\langle\nabla_\theta\mathcal{L}_{\mathrm{util}},\nabla_\theta\mathcal{L}_{\mathrm{priv}}\rangle\approx0$ (Section~\ref{sec:ortho}), changing the privacy loss shifts the diagnostic loss only through a first-order term that vanishes, so the two objectives behave as independent directions.
The privacy--utility trade-off is therefore escapable rather than fundamental, and REAN can suppress biometric information without sacrificing diagnosis.

We maximize correct-class cross-entropy rather than output entropy.
Maximizing correct-class cross-entropy directly reduces the true-class probability, whereas entropy can stay high even when the true class remains top-ranked.
One privacy-loss weight controls the bundled $\mathrm{CE}_{\mathrm{bio}}$, matching the single near-orthogonal biometric direction of Section~\ref{sec:ortho}.
Each term carries a nonnegative weight, selected on validation data by Bayesian optimization with a Tree-structured Parzen Estimator~\cite{tpe} that approaches the ideal corner of privacy at chance, utility maximal, and distortion minimal.
The selected weights for utility, privacy, PRD, and $L_1$ are $(1.51, 0.153, 0.323, 1.46)$, and nearby high-scoring configurations preserve the same privacy--utility--distortion pattern.

\subsection{Reconstruction with a 1-D U-Net}
\label{sec:unet}
REAN realizes the objective as a 1-D U-Net that reconstructs the ECG by adding a bounded, input-conditioned residual.
Given a window $\mathbf{x}\in\mathbb{R}^{L}$, REAN outputs
\begin{equation}
  \hat{\mathbf{x}} = \mathbf{x} + \tanh\!\big(\delta_\theta(\mathbf{x})\big)\cdot
  \varepsilon_{\max}.
  \label{eq:output}
\end{equation}
Here $\delta_\theta$ is the residual predictor, implemented as a U-Net with an encoder--decoder structure, strided convolutions, skip connections, base width $C{=}32$, and four resolution levels~\cite{unet}.
The $\tanh$ bounds the residual to $[-\varepsilon_{\max},\varepsilon_{\max}]$ and keeps $\hat{\mathbf{x}}$ within the normalized signal range.

A learned residual is what lets REAN follow the privacy direction rather than move blindly.
Section~\ref{sec:ortho} shows that a privacy direction exists that suppresses biometric information while leaving diagnosis intact.
Blind noise moves the signal in every direction at once, so it disturbs diagnosis and does not exploit this direction, whereas a residual conditioned on the input can align with it.
REAN therefore reconstructs the signal with a U-Net rather than adding undirected noise.

The amplitude bound is selected to suppress protected attributes without leaving the diagnostic-quality range.
We set $\varepsilon_{\max}{=}0.08$, the smallest validation-selected value at which biometric inference reaches random-chance levels while PRD stays in the diagnostic-quality band~\cite{zigel2000}.
REAN needs no per-signal detection or recalibration at inference; it conditions the perturbation on the input and produces it in one pass.

\begin{table*}[t]
\centering
\caption{Public databases in the benchmark (all from PhysioNet, resampled to 250\,Hz, 8-s windows). Subjects is the number of subjects per database (186 total; 184 demographically labeled); age and gender are as reported by each source database.}
\label{tab:data}
\setlength{\tabcolsep}{6pt}
\begin{tabular}{lccccl}
\toprule
Database & Subjects & Age Range & Gender (M/F) & Sampling rate (Hz) & Health condition \\
\midrule
MIT-BIH Arrhythmia~\cite{mitbih}   & 47  & 23--89 & 25/22 & 360 & mixed arrhythmia \\
MIT-BIH Long-Term~\cite{ltdb}      & 7   & 46--88 & 6/1   & 128 & long-term general monitoring \\
INCART~\cite{incartdb,physionet}   & 32  & 18--80 & 17/15 & 257 & coronary artery disease / arrhythmia \\
SHDB-AF~\cite{shdbaf}              & 100 & 31--87 & 55/45 & 200 & atrial fibrillation \\
\midrule
\textbf{Total} & \textbf{186} & & & $\to$\,250 & 1.16\,M windows \\
\bottomrule
\end{tabular}
\end{table*}

\textbf{Training and inference.} Training updates only REAN, and inference uses only REAN.
The three frozen biometric classifiers are ECGViT models and the frozen diagnostic classifier is a 1-D ResNet~\cite{resnet}.
REAN is trained with AdamW ($\mathrm{lr}{=}10^{-3}$, $\mathrm{wd}{=}10^{-4}$)~\cite{adamw}, cosine annealing~\cite{sgdr}, batch size 64, and 100 epochs on a balanced 30{,}000-window subsample; the test set is never used for selection.
At inference, a single U-Net forward pass anonymizes each window independent of cohort statistics, which is why REAN is faster than per-sample methods (Section~\ref{sec:ablation}).


\section{Experiments}
\label{sec:exp}

\subsection{Setup}
\label{sec:setup}
\textbf{Datasets.} We evaluate REAN on a four-database PhysioNet benchmark with 186 identities, of which 184 carry demographic labels (Table~\ref{tab:data}).
We resample all recordings to 250\,Hz and segment them into 8-second (2000-sample) windows with per-window min-max normalization, producing 1{,}163{,}983 windows.
Arrhythmia labels are Normal (77.3\%), AFIB (19.0\%), PVC (3.4\%), and SVE (0.3\%).
All four models use the same per-participant temporal split (70/15/15) without segment overlap, so every subject appears in train, validation, and test partitions while time segments do not overlap.
Identity is evaluated as closed-set re-identification with enrollment recordings, and gender and age as attribute inference from unseen recordings of cohort patients.
A shared 500-window held-out set keeps the baseline comparison fair while bounding REACT's evaluation cost; the set is gender-stratified (250/250) and preserves the arrhythmia distribution.
Because REAN is amortized, we also evaluate Clean and REAN on a larger 20{,}000-window subset of the 177{,}329-window test split to test whether the REAN trend persists at scale.

\textbf{Frozen classifiers.} Strong frozen classifiers define both the training signal and the privacy threat model.
The three biometric classifiers are ECGViT models trained following the TransECG protocol~\cite{transecg} (patch 20, embedding 240, depth 6, 6 heads, MLP 128, stochastic depth survival 0.8), and the diagnostic classifier is a 1-D ResNet.
On clean data these models reach identity 0.964, gender 0.974, age 0.976, and arrhythmia 0.978.
This strength matters because a weak biometric classifier would overstate privacy.

\textbf{Baselines.} We compare REAN with seven baselines under identical data, models, and evaluation conditions.
The baselines are full-signal Gaussian and Laplacian noise, REACT~\cite{react}, Lee et al.~\cite{leeetal}, PP-VAE~\cite{ppvae}, PrivECG~\cite{privecg}, and TransECG~\cite{transecg}.
REACT is a reinforcement-learning method that re-optimizes noise per record with PPO~\cite{ppo}.
Lee et al.\ finds identity-related feature regions with an attention mapper and perturbs a feature representation; because its native output is a feature rather than a signal, we preserve its attention-difference weighted noise mechanism and add a raw-to-feature encoder and feature-to-raw decoder so the input and output are raw waveforms.
PP-VAE targets demographic privacy in a VAE latent space and does not address individual re-identification, and PrivECG generates an anonymized ECG with a GAN.
TransECG is a re-identification-risk analyzer rather than an anonymizer, so we instantiate its anonymization claim as a segment-noise baseline that injects Gaussian noise into the union of the QRS and P--R segments its attention ranks as most identifying.
All methods use the same four frozen classifiers; learned baselines follow their paper's optimizer and schedule, while TransECG, Gaussian, and Laplacian are training-free.

\textbf{Metrics.} The benchmark measures privacy, diagnostic utility, distortion, and speed.
Privacy is ReID accuracy, raw gender accuracy, and age accuracy, with random performance $\approx$0.005/0.500/0.200 ($1/186$, $1/2$, $1/5$); lower is better for ReID and age, and for binary gender the privacy-relevant quantity is the distance from 0.5, because a below-chance classifier is invertible by flipping its decision.
Diagnostic utility (higher better) is arrhythmia accuracy and macro-AUROC, and macro-AUROC is preferred because SVE is only 0.3\% of the data and weighted averaging would hide rare-class failure.
Distortion (lower better) is PRD, which the compression literature considers ``very good'' below 9\%~\cite{zigel2000}.
Speed is anonymization time per window.
An ideal method meets all four requirements at once, as summarized by Table~\ref{tab:main} and Fig.~\ref{fig:pareto}.

\subsection{Escaping the privacy--utility trade-off}
\label{sec:main}
\begin{figure}[t]
\centering
\includegraphics[width=\columnwidth]{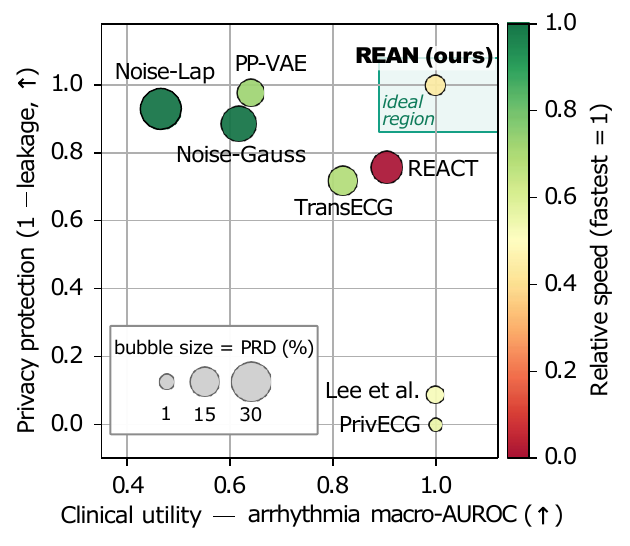}
\caption{Privacy--utility--speed. $x$: arrhythmia macro-AUROC ($\uparrow$); $y$: privacy protection ($1-$normalized leakage, $\uparrow$); bubble size: PRD (distortion); color: relative speed (fastest $=1$; green fast, red slow). This aggregate uses the raw leakage metrics in Table~\ref{tab:main}; the binary-gender inversion caveat is reported separately in the table caption. REAN reaches the high-utility, high-protection corner while staying in the fast amortized band.}
\label{fig:pareto}
\end{figure}

REAN is the only method that suppresses biometric leakage while keeping diagnosis at the clean level (Table~\ref{tab:main}, Fig.~\ref{fig:pareto}).
Against the pretrained attacker classifiers, REAN drives ReID and age accuracy to at or below chance and moves the raw gender prediction far from the clean response.\footnote{Gender is binary, so the privacy-relevant quantity is the distance from chance ($0.5$), not whether raw accuracy is above or below $0.5$; a below-chance classifier can be inverted by flipping the decision.
We therefore report gender as raw binary accuracy for transparency, do not mark a gender winner in Table~\ref{tab:main}, and treat retraining-aware recovery as the stronger gender test (Table~\ref{tab:advtrain}).}
REAN keeps arrhythmia macro-AUROC at the clean level.
The point estimate is slightly higher than clean (0.9991 vs.\ 0.9982), but a paired subject-clustered bootstrap gives a 95\% confidence interval of $[-0.0006,\,0.0034]$ for REAN--Clean, so we read the difference as statistically indistinguishable rather than as an improvement.

The orthogonality is realized in training.
From clean to anonymized signals the diagnostic cross-entropy is essentially unchanged ($0.075\!\to\!0.076$), while the biometric cross-entropies rise by over two orders of magnitude (ReID $0.23\!\to\!42.6$, gender $0.10\!\to\!10.5$, age $0.09\!\to\!16.4$).
REAN drives the attackers' loss up steeply while paying almost no diagnostic cost, the direct signature of optimizing a biometric objective whose gradient is orthogonal to the diagnostic one.

\begin{figure}[t]
\centering
\includegraphics[width=\columnwidth]{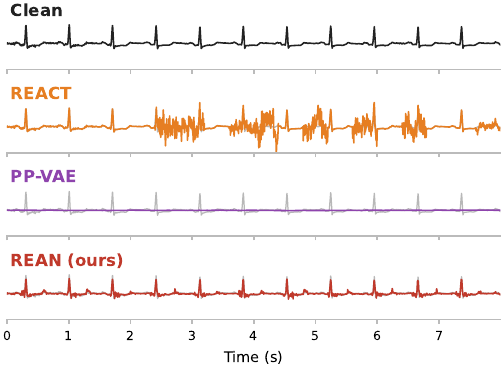}
\caption{Waveforms on a representative 8\,s test window. Each panel overlays the anonymized trace (color) on the clean signal (gray). REACT injects visible per-sample noise and PP-VAE collapses the morphology to a near-flat trace, whereas REAN closely tracks the clean waveform. Per-method distortion (PRD) is reported in Table~\ref{tab:main}.}
\label{fig:waveform}
\end{figure}

The baselines each trade off at least one axis.
PP-VAE, random noise, TransECG, and REACT reduce leakage only by sacrificing utility or distortion, and Fig.~\ref{fig:waveform} contrasts these distortions with REAN.
PrivECG and Lee et al.~\cite{leeetal} preserve utility or waveform similarity but leave identity recoverable (0.966 and 0.882).
The Lee et al.\ noise reduces its own attention-mapper identity head, yet the output stays re-identifiable by an independent biometric classifier, which shows that anonymizing against oneself does not transfer to an independent attacker.

The REAN trend persists at larger scale, so this REAN-only check supports the shared 500-window comparison rather than replacing it.
On a single 20{,}000-window subset of the test split, Clean and REAN macro-AUROC are 0.9867 and 0.9945, and REAN attains ReID 0.00025, gender 0.488, age 0.082, and PRD 6.77\%.

\begin{table}[t]
\centering
\caption{Main results on the shared 500-window evaluation set. Bold marks the best privacy and utility values among privacy-protective methods, except for Gender; PRD has no winner marker. Gender is raw binary accuracy, so below-chance values are invertible and must be read by distance from 0.5. REAN's AUROC is indistinguishable from clean (paired subject-clustered 95\% CI: $[-0.0006,\,0.0034]$).}
\label{tab:main}
\small
\setlength{\tabcolsep}{3pt}
\begin{tabular}{lcccccc}
\toprule
\multirow{2}{*}{Method}
  & \multicolumn{3}{c}{Privacy}
  & \multicolumn{2}{c}{Utility ($\uparrow$)}
  & Distort. \\
\cmidrule(lr){2-4}\cmidrule(lr){5-6}\cmidrule(lr){7-7}
  & ReID$\downarrow$ & Gender & Age$\downarrow$ & Arr & AUROC & PRD \\
\midrule
Clean (no def.)           & 0.964 & 0.974 & 0.976 & 0.978 & 0.9982 & 0.00 \\
Random chance             & $\approx$.005 & 0.500 & 0.200 & --- & --- & --- \\
\midrule
Noise-Gauss               & 0.058 & 0.572 & 0.304 & 0.770 & 0.6161 & 24.00 \\
Noise-Lap                 & 0.032 & 0.568 & 0.232 & 0.770 & 0.4653 & 33.70 \\
REACT~\cite{react}        & 0.210 & 0.584 & 0.462 & 0.818 & 0.9034 & 18.20 \\
TransECG                  & 0.250 & 0.610 & 0.478 & 0.754 & 0.8185 & 14.80 \\
PP-VAE~\cite{ppvae}       & 0.004 & 0.452 & 0.252 & 0.110 & 0.6389 & 11.91 \\
PrivECG~\cite{privecg}    & 0.966 & 0.976 & 0.978 & 0.980 & 0.9981 & 0.30 \\
Lee et al.~\cite{leeetal} & 0.882 & 0.912 & 0.940 & 0.968 & 0.9975 & 3.17 \\
\midrule
\textbf{REAN (ours)}      & \textbf{0.000} & 0.142 & \textbf{0.044} & \textbf{0.974} & \textbf{0.9991} & 5.56 \\
\bottomrule
\end{tabular}
\end{table}

\subsection{Privacy robustness}
\label{sec:privacy}
REAN's re-identification protection holds under adaptive and unseen attacks.
We test three threats stronger than the pretrained attackers of Section~\ref{sec:main}.

\textbf{Input purification.} An attacker who receives the anonymized signal $\hat{\mathbf{x}}$ cannot restore identity by denoising.
We apply Butterworth 40\,Hz low-pass filtering, db4 wavelet soft thresholding, median filtering, Savitzky--Golay filtering, and a learned denoising autoencoder trained on clean ECG~\cite{butterworth1930,donoho1995,savitzky1964,dae}, applied only to REAN's output.
After purification, ReID stays at or near chance across all purifiers, while gender and age are partially recovered only by the stronger learned restorers (Table~\ref{tab:purify}).
If REAN's residual $\delta$ were random high-frequency noise, purification would restore identity, so this result shows that $\delta$ is not removed by standard signal denoising.

\textbf{Unseen attacker architectures.} REAN's protection transfers to attackers it never trained against.
REAN is trained against ECGViT biometric classifiers, so we re-attack the anonymized signal with 1-D ResNet, InceptionTime~\cite{inceptiontime}, and CNN-LSTM.
Identity stays substantially reduced relative to clean while gender and age are partially recovered (gender 0.43--0.60, age 0.28--0.42), so ReID protection is not overfit to the trained architecture (Table~\ref{tab:transfer}) and is consistent with the backbone-invariant orthogonality of Section~\ref{sec:ortho}.

\textbf{Retraining-aware adversary.} Residual biometric structure remains learnable to an adversary that retrains on anonymized data.
Recovery grows with the anonymized-data share: ReID rises from 0.000 to 0.590 at a 20\% mix, 0.726 at 50\%, and 0.886 at 100\% (Table~\ref{tab:advtrain}).
The 100\% setting is a worst case that requires a fully labeled anonymized corpus, and read as an upper bound it shows that a deterministic map can leave residual biometric structure.
This is a structural property of any deterministic anonymizer, which we discuss in Section~\ref{sec:discussion}.

\begin{table}[t]
\centering
\caption{Attribute recovery after applying a purifier to REAN's output. Lower ReID/Age is more robust; Gender follows the raw-binary caveat in Table~\ref{tab:main}. ``PRD$\to$clean'' is distance from the original.}
\label{tab:purify}
\setlength{\tabcolsep}{5pt}
\begin{tabular}{lcccc}
\toprule
Purifier & ReID & Gender & Age & PRD$\to$clean \\
\midrule
Clean (recovery bound) & 0.964 & 0.974 & 0.976 & 0.00 \\
Random chance          & $\approx$.005 & 0.500 & 0.200 & --- \\
REAN (no purify)       & 0.000 & 0.142 & 0.044 & 5.56 \\
\midrule
Low-pass 40\,Hz        & 0.000 & 0.174 & 0.046 & 5.14 \\
Wavelet (db4)          & 0.000 & 0.166 & 0.048 & 5.12 \\
Median filter          & 0.000 & 0.176 & 0.058 & 5.19 \\
Savitzky--Golay        & 0.012 & 0.474 & 0.198 & 5.42 \\
DAE (learned)          & 0.036 & 0.560 & 0.398 & 6.80 \\
\bottomrule
\end{tabular}
\end{table}

\begin{table}[t]
\centering
\caption{Attacking the anonymized signal with unseen architectures. Gender follows the raw-binary caveat in Table~\ref{tab:main}.}
\label{tab:transfer}
\setlength{\tabcolsep}{6pt}
\begin{tabular}{lccc}
\toprule
Attacker & ReID & Gender & Age \\
\midrule
Random         & $\approx$.005 & 0.500 & 0.200 \\
\midrule
ResNet1D       & 0.092 & 0.580 & 0.422 \\
InceptionTime  & 0.026 & 0.432 & 0.276 \\
CNN-LSTM       & 0.050 & 0.598 & 0.288 \\
\bottomrule
\end{tabular}
\end{table}

\begin{table}[t]
\centering
\caption{Retraining-aware attack: the attacker retrains a biometric classifier as the share of anonymized data grows. Gender follows the raw-binary caveat in Table~\ref{tab:main}.}
\label{tab:advtrain}
\setlength{\tabcolsep}{6pt}
\begin{tabular}{lccc}
\toprule
Training mix & ReID & Gender & Age \\
\midrule
20\% anon + 80\% clean & 0.590 & 0.614 & 0.672 \\
50\% anon + 50\% clean & 0.726 & 0.818 & 0.702 \\
100\% anon             & 0.886 & 0.938 & 0.840 \\
\bottomrule
\end{tabular}
\end{table}

\subsection{Diagnostic utility}
\label{sec:utility}
REAN preserves diagnostic utility across arrhythmia classes and at natural prevalence.
On the shared comparison set, per-class AUROC stays near clean for all four classes, and macro-AUROC (the mean of the four one-vs-rest class AUROCs) matches Table~\ref{tab:main} (Table~\ref{tab:perclass}).

We further validate the rare SVE class beyond the shared 500-window set.
On a natural-prevalence 20{,}000-window pool (183 subjects, 54 SVE windows), 30 stratified resamplings show the REAN--Clean macro-AUROC estimate stabilizing as the sample grows, reaching 0.9972 for REAN against 0.9956 for clean.
An SVE-enriched balanced evaluation using all 469 SVE windows and 469 windows from each other class (1{,}876 windows, 162 subjects) gives the same picture, with macro-AUROC 0.9977 against clean 0.9930 and SVE AUROC 0.9977 against clean 0.9842.
These checks confirm that the utility result is not a sample-size or rare-class artifact.

\begin{table}[t]
\centering
\caption{Per-class arrhythmia AUROC (SVE is 0.3\% of the test split). REAN (ours) last; Macro is the mean over the four classes.}
\label{tab:perclass}
\setlength{\tabcolsep}{4pt}
\begin{tabular}{lccccc}
\toprule
Method & Normal & PVC & SVE & AFIB & Macro \\
\midrule
Clean                     & 0.9963 & 1.0000 & 1.0000 & 0.9963 & 0.9982 \\
Noise-Gauss               & 0.5601 & 0.7356 & 0.6473 & 0.5212 & 0.6161 \\
Noise-Lap                 & 0.4681 & 0.6088 & 0.3006 & 0.4839 & 0.4653 \\
REACT~\cite{react}        & 0.8804 & 0.9885 & 0.8297 & 0.9149 & 0.9034 \\
TransECG                  & 0.7605 & 0.8679 & 0.9559 & 0.6895 & 0.8185 \\
PP-VAE~\cite{ppvae}       & 0.5372 & 0.7651 & 0.6012 & 0.6521 & 0.6389 \\
PrivECG~\cite{privecg}    & 0.9962 & 1.0000 & 1.0000 & 0.9961 & 0.9981 \\
Lee et al.~\cite{leeetal} & 0.9943 & 0.9988 & 1.0000 & 0.9970 & 0.9975 \\
\midrule
\textbf{REAN (ours)}      & 0.9978 & 0.9997 & 1.0000 & 0.9990 & 0.9991 \\
\bottomrule
\end{tabular}
\end{table}

\subsection{Ablation and speed}
\label{sec:ablation}
No single objective term is sufficient for the privacy--utility--distortion balance (Table~\ref{tab:ablation}).
Turning the three bundled terms (U, P, D) on and off separates their roles.
Single-term objectives each fail, since U leaks identity, P harms diagnosis, and D preserves the input without protection.
Pairwise objectives remove one failure but leave another, since U\,P distorts, P\,D hurts diagnosis, and U\,D leaks.
Only the full objective (U\,P\,D) satisfies every requirement at once, reaching identity at chance, arrhythmia 0.974, and PRD 5.56, and it attains the smallest distance to the ideal corner ($d{=}0.28$).

\textbf{Speed.} As a secondary benefit, REAN is fast because it anonymizes in a single forward pass.
The speed comes from amortization, not from the orthogonality, since REAN runs the U-Net once per window with no per-record optimization.
We measure GPU latency on an RTX 4090 (PyTorch 2.11.0, CUDA 13.0), timing anonymized-signal generation on the same 256-window batch with GPU synchronization.
REAN takes 0.117\,ms/window, within the amortized band below 1\,ms/window, whereas REACT takes 12.73\,ms/window ($108.8\times$ REAN) because it re-optimizes per record.
The other fast methods each fail a different requirement.
PrivECG leaves identity recoverable, PP-VAE collapses utility, TransECG damages utility, and Lee et al.\ does not return a reusable raw waveform.

\begin{table}[t]
\centering
\caption{Objective ablation (U=Utility, P=Privacy, D=Distortion). REAN (UPD) last. $d$ is the distance to the ideal corner (mean privacy leakage $+$ utility gap $+$ PRD$/20$; lower better). Gender follows the raw-binary caveat in Table~\ref{tab:main}.}
\label{tab:ablation}
\setlength{\tabcolsep}{3.4pt}
\begin{tabular}{lcccccc}
\toprule
Terms & ReID & Gender & Age & Arr-Acc & PRD\% & $d\downarrow$ \\
\midrule
U          & 0.704 & 0.830 & 0.842 & 0.990 & 10.27 & 1.29 \\
P          & 0.000 & 0.094 & 0.104 & 0.730 & 15.45 & 1.02 \\
D          & 0.964 & 0.974 & 0.976 & 0.978 & 0.00  & 1.00 \\
U\,P       & 0.000 & 0.096 & 0.074 & 0.962 & 15.59 & 0.80 \\
U\,D       & 0.964 & 0.974 & 0.976 & 0.978 & 0.00  & 1.00 \\
P\,D       & 0.000 & 0.146 & 0.016 & 0.810 & 5.65  & 0.45 \\
\midrule
\textbf{U\,P\,D}    & 0.000 & 0.142 & 0.044 & 0.974 & 5.56 & \textbf{0.28} \\
\bottomrule
\end{tabular}
\end{table}

\section{Related Work}
\label{sec:related}
\textbf{ECG-based biometrics.} ECG biometrics make signal-level defenses necessary.
Fiducial and deep models identify people and infer attributes from ECG signals~\cite{ecgunveiled,transecg}, and partial side knowledge can re-identify public ECG records~\cite{linkageattacks,deanon}.
REAN therefore evaluates against strong biometric classifiers.

\textbf{Anonymization defenses} address complementary deployment settings.
Feature- and latent-space transforms are attractive when the downstream extractor is fixed, because they can be fast and compact.
Lee et al.~\cite{leeetal} perturbs identity-related features, but its native output is extractor-tied and stays re-identifiable after our raw-waveform bridge.
PP-VAE~\cite{ppvae} targets demographic privacy in a latent representation, but its representative-beat output does not preserve the rhythm information the arrhythmia task needs.

Signal-space defenses return deployable waveforms, but they trade off speed, privacy, and diagnostic fidelity in different ways.
PrivECG~\cite{privecg} keeps high signal fidelity but leaves strong biometric classifiers accurate in our benchmark.
Iterative methods such as REACT~\cite{react} optimize protection per record, but their per-sample optimization makes large-scale deployment slow.
Full-signal noise and differential privacy~\cite{dpecg} are fast, but perturb diagnostic morphology without using the diagnostic--biometric geometry.

\textbf{Position of REAN.} REAN combines reusable raw ECG output, a multi-attribute privacy loss, and a geometric basis for preserving diagnosis while suppressing biometrics, with single-pass speed as a secondary benefit.

\section{Discussion}
\label{sec:discussion}
\textbf{Why orthogonality exists.} Orthogonality likely arises because identity and arrhythmia rely on different ECG sub-structures.
Identity is carried mostly by the person-specific shape of the QRS complex, while arrhythmia diagnosis depends on rhythm and broader P and T morphology.
These different signal dependencies make the direction that blurs identity nearly orthogonal to the direction that determines diagnosis, and REAN learns this structure from frozen-model gradients without an explicit rule.

\textbf{Limitations.} REAN's central privacy limitation is deterministic leakage under a retraining-aware adversary.
For a deterministic map $\mathbf{x}\!\mapsto\!\hat{\mathbf{x}}$ and a biometric attribute $y_a$, the data-processing inequality~\cite{coverthomas} gives $I(\hat{\mathbf{x}};y_a)\le I(\mathbf{x};y_a)$ rather than equality, so residual biometric structure can remain and be recovered by retraining (Section~\ref{sec:privacy}).
Formal control of this leakage requires stochastic mechanisms such as differential privacy~\cite{dpecg,dwork2006} or a variational bottleneck~\cite{vib}, so REAN is best viewed as practical signal-level obfuscation rather than formal irreversible anonymization.
A second limitation is evaluation scope, since our evaluation uses single-lead ECG with a within-subject temporal split, leaving multi-lead and 12-lead recordings, unseen-subject deployment, and non-arrhythmia clinical domains to future work.

\section{Conclusion}
\label{sec:conclusion}
We proposed REAN, a raw ECG waveform anonymizer for privacy-preserving ECG sharing.
The central obstacle was the privacy--utility trade-off, since anonymization must suppress biometric information without destroying diagnostic morphology.
We showed that this trade-off is escapable because the diagnostic and biometric objectives have near-orthogonal gradients, and REAN turned this geometry into a lightweight learned transform.
Using frozen diagnostic and biometric classifiers only during training, REAN learned a single-pass U-Net anonymizer that suppresses biometric leakage while preserving diagnosis.
On four PhysioNet databases, REAN drove re-identification to chance, kept arrhythmia diagnosis at the clean level, and transferred re-identification protection to unseen attacker architectures, at over an order of magnitude less time than per-sample optimization.
The remaining challenge is retraining-aware leakage, and future work should add stochastic mechanisms and extend evaluation to multi-lead and multi-domain settings.

\bibliographystyle{ieeetr}
\bibliography{refs}

@inproceedings{ecgunveiled,
  title={Ecg unveiled: Analysis of client re-identification risks in real-world ecg datasets},
  author={Wang, Ziyu and Kanduri, Anil and Aqajari, Seyed Amir Hossein and Jafarlou, Salar and Mousavi, Sanaz R and Liljeberg, Pasi and Malik, Shaista and Rahmani, Amir M},
  booktitle={2024 IEEE 20th International Conference on Body Sensor Networks (BSN)},
  pages={1--4},
  year={2024},
  organization={IEEE}
}

@article{transecg,
  title={Transecg: Leveraging transformers for explainable ecg re-identification risk analysis},
  author={Wang, Ziyu and Khatibi, Elahe and Kazemi, Kianoosh and Azimi, Iman and Mousavi, Sanaz and Malik, Shaista and Rahmani, Amir M},
  journal={arXiv preprint arXiv:2503.13495},
  year={2025}
}

@inproceedings{react,
  title={React: Reinforcement learning-based adaptive ecg anonymization and privacy threat mitigation},
  author={Datta, Aniruddha and Bhattacharyya, Tamonash and Khatibi, Elahe and Seth, Agasthya and Wang, Ziyu and Mousavi, Sanaz Rahimi and Rahmani, Amir M and Firouzi, Farshad and Chakrabarty, Krishnendu},
  booktitle={2025 IEEE International Conference on Omni-layer Intelligent Systems (COINS)},
  pages={1--8},
  year={2025},
  organization={IEEE}
}

@article{ppvae,
  title={AI-enabled privacy-preserving cardiac diagnostics via electrocardiograms},
  author={Shishir, Fairuz Shadmani and Harvey, Christopher J and Gupta, Amulya and Noheria, Amit and Shomaji, Sumaiya},
  journal={Scientific Reports},
  year={2026},
  publisher={Nature Publishing Group UK London}
}

@article{leeetal,
  title={Privacy-preserving ECG data collection for arrhythmia classification},
  author={Lee, Hyubjin and Kim, Minsoo and Chung, Yon Dohn},
  journal={Biomedical Signal Processing and Control},
  volume={112},
  pages={108374},
  year={2026},
  publisher={Elsevier}
}

@inproceedings{linkageattacks,
  title={Linkage Attacks Expose Identity Risks in Public ECG Data Sharing},
  author={Wang, Ziyu and Khatibi, Elahe and Firouzi, Farshad and Mousavi, Sanaz Rahimi and Chakrabarty, Krishnendu and Rahmani, Amir M},
  booktitle={2025 47th Annual International Conference of the IEEE Engineering in Medicine and Biology Society (EMBC)},
  pages={1--7},
  year={2025},
  organization={IEEE}
}

@inproceedings{deanon,
  title={ECG De-Anonymization: Real-World Risks and a Privacy-by-Design Mitigation Strategy},
  author={Aguelal, Hamza and Palmieri, Paolo},
  booktitle={2025 IEEE 38th International Symposium on Computer-Based Medical Systems (CBMS)},
  pages={449--456},
  year={2025},
  organization={IEEE}
}

@inproceedings{privecg,
  title={Privecg: Generating private ecg for end-to-end anonymization},
  author={Nolin-Lapalme, Alexis and Avram, Robert and Julie, Hussin},
  booktitle={Machine Learning for Healthcare Conference},
  pages={509--528},
  year={2023},
  organization={PMLR}
}

@article{dpecg,
  title={Privacy-preserving ecg data analysis with differential privacy: A literature review and a case study},
  author={Ghazarian, Arin and Zheng, Jianwei and Rakovski, Cyril},
  journal={arXiv preprint arXiv:2406.13880},
  year={2024}
}

@article{zigel2000,
  title={The weighted diagnostic distortion (WDD) measure for ECG signal compression},
  author={Zigel, Yaniv and Cohen, Arnon and Katz, Amos},
  journal={IEEE transactions on biomedical engineering},
  volume={47},
  number={11},
  pages={1422--1430},
  year={2000},
  publisher={IEEE}
}

@article{mitbih,
  title={The impact of the MIT-BIH arrhythmia database},
  author={Moody, George B and Mark, Roger G},
  journal={IEEE engineering in medicine and biology magazine},
  volume={20},
  number={3},
  pages={45--50},
  year={2001},
  publisher={IEEE}
}

@article{physionet,
  title={PhysioBank, PhysioToolkit, and PhysioNet: components of a new research resource for complex physiologic signals},
  author={Goldberger, Ary L and Amaral, Luis AN and Glass, Leon and Hausdorff, Jeffrey M and Ivanov, Plamen Ch and Mark, Roger G and Mietus, Joseph E and Moody, George B and Peng, Chung-Kang and Stanley, H Eugene},
  journal={circulation},
  volume={101},
  number={23},
  pages={e215--e220},
  year={2000},
  publisher={Lippincott Williams \& Wilkins}
}

@article{ltdb,
  author  = {Goldberger, Ary L. and Amaral, Luis A. N. and Glass, Leon and Hausdorff, Jeffrey M. and Ivanov, Plamen Ch. and Mark, Roger G. and Mietus, Joseph E. and Moody, George B. and Peng, Chung-Kang and Stanley, H. Eugene},
  title   = {PhysioBank, PhysioToolkit, and PhysioNet: Components of a New Research Resource for Complex Physiologic Signals},
  journal = {Circulation},
  year    = {2000},
  volume  = {101},
  number  = {23},
  pages   = {e215--e220},
  note    = {RRID:SCR\_007345}
}

@article{incartdb,
  title={St Petersburg INCART 12-Lead Arrhythmia Database. 2008},
  author={Tihonenko, V and Khaustov, A and Ivanov, S and Rivin, A and Yakushenko, E},
  year={2008},
  journal={PhysioBank PhysioToolkit and PhysioNet}
}

@article{shdbaf,
  title={SHDB-AF: a Japanese Holter ECG database of atrial fibrillation},
  author={Tsutsui, Kenta and Brimer, Shany Biton and Ben-Moshe, Noam and Sellal, Jean Marc and Oster, Julien and Mori, Hitoshi and Ikeda, Yoshifumi and Arai, Takahide and Nakano, Shintaro and Kato, Ritsushi and others},
  journal={Scientific data},
  volume={12},
  number={1},
  pages={454},
  year={2025},
  publisher={Nature Publishing Group UK London}
}

@article{inceptiontime,
  title={Inceptiontime: Finding alexnet for time series classification},
  author={Ismail Fawaz, Hassan and Lucas, Benjamin and Forestier, Germain and Pelletier, Charlotte and Schmidt, Daniel F and Weber, Jonathan and Webb, Geoffrey I and Idoumghar, Lhassane and Muller, Pierre-Alain and Petitjean, Fran{\c{c}}ois},
  journal={Data mining and knowledge discovery},
  volume={34},
  number={6},
  pages={1936--1962},
  year={2020},
  publisher={Springer}
}

@inproceedings{unet,
  title={U-net: Convolutional networks for biomedical image segmentation},
  author={Ronneberger, Olaf and Fischer, Philipp and Brox, Thomas},
  booktitle={International Conference on Medical image computing and computer-assisted intervention},
  pages={234--241},
  year={2015},
  organization={Springer}
}

@inproceedings{resnet,
  title={Deep residual learning for image recognition},
  author={He, Kaiming and Zhang, Xiangyu and Ren, Shaoqing and Sun, Jian},
  booktitle={Proceedings of the IEEE conference on computer vision and pattern recognition},
  pages={770--778},
  year={2016}
}

@article{tpe,
  title={Algorithms for hyper-parameter optimization},
  author={Bergstra, James and Bardenet, R{\'e}mi and Bengio, Yoshua and K{\'e}gl, Bal{\'a}zs},
  journal={Advances in neural information processing systems},
  volume={24},
  year={2011}
}

@article{adamw,
  title={Decoupled weight decay regularization},
  author={Loshchilov, Ilya and Hutter, Frank},
  journal={arXiv preprint arXiv:1711.05101},
  year={2017}
}

@inproceedings{sgdr,
  author    = {I. Loshchilov and F. Hutter},
  title     = {{SGDR}: Stochastic Gradient Descent with Warm Restarts},
  booktitle = {Proc. Int. Conf. Learning Representations (ICLR)},
  year      = {2017},
  note      = {arXiv:1608.03983},
}

@article{ppo,
  title={Proximal policy optimization algorithms},
  author={Schulman, John and Wolski, Filip and Dhariwal, Prafulla and Radford, Alec and Klimov, Oleg},
  journal={arXiv preprint arXiv:1707.06347},
  year={2017}
}

@article{butterworth1930,
  title={On the theory of filter amplifiers},
  author={Butterworth, Stephen and others},
  journal={Wireless Engineer},
  volume={7},
  number={6},
  pages={536--541},
  year={1930}
}

@article{donoho1995,
  title={De-noising by soft-thresholding},
  author={Donoho, David L},
  journal={IEEE transactions on information theory},
  volume={41},
  number={3},
  pages={613--627},
  year={1995},
  publisher={IEEE}
}

@article{savitzky1964,
  title={Smoothing and differentiation of data by simplified least squares procedures.},
  author={Savitzky, Abraham and Golay, Marcel JE},
  journal={Analytical chemistry},
  volume={36},
  number={8},
  pages={1627--1639},
  year={1964},
  publisher={ACS Publications}
}

@inproceedings{dae,
  title={Extracting and composing robust features with denoising autoencoders},
  author={Vincent, Pascal and Larochelle, Hugo and Bengio, Yoshua and Manzagol, Pierre-Antoine},
  booktitle={Proceedings of the 25th international conference on Machine learning},
  pages={1096--1103},
  year={2008}
}

@book{coverthomas,
  title={Elements of information theory (wiley series in telecommunications and signal processing)},
  author={Cover, Thomas M and Thomas, Joy A},
  year={2006},
  publisher={Wiley-interscience}
}

@inproceedings{dwork2006,
  title={Calibrating noise to sensitivity in private data analysis},
  author={Dwork, Cynthia and McSherry, Frank and Nissim, Kobbi and Smith, Adam},
  booktitle={Theory of cryptography conference},
  pages={265--284},
  year={2006},
  organization={Springer}
}

@inproceedings{vib,
  author    = {A. A. Alemi and I. Fischer and J. V. Dillon and K. Murphy},
  title     = {Deep Variational Information Bottleneck},
  booktitle = {Proc. Int. Conf. Learning Representations (ICLR)},
  year      = {2017},
  note      = {arXiv:1612.00410},
}

\end{document}